\documentclass{PoS}

\newcommand{\dd}{{\rm{d}}}
\newcommand{\Fin}{F_{\mbox{\tiny{in}}}}
\newcommand{\Ffin}{F_{\mbox{\tiny{fin}}}}
\newcommand{\lambdain}{\lambda_{\mbox{\tiny{in}}}}
\newcommand{\lambdafin}{\lambda_{\mbox{\tiny{fin}}}}
\newcommand{\Zp}{Z_{\mbox{\tiny{p}}}}
\newcommand{\Za}{Z_{\mbox{\tiny{a}}}}
\newcommand{\SU}{\mathrm{SU}}
\newcommand{\betagauge}{\beta_{\mbox{\tiny{g}}}}
\newcommand{\betagaugein}{\beta_{\mbox{\tiny{g,i}}}}
\newcommand{\betagaugefin}{\beta_{\mbox{\tiny{g,f}}}}
\newcommand{\nr}{n_{\mbox{\tiny{r}}}}

\title{Applications of Jarzynski's relation in lattice gauge theories}

\ShortTitle{Applications of Jarzynski's relation in lattice gauge theories}

\author{\speaker{Alessandro Nada}$^a$, Michele~Caselle,$^a$ Gianluca~Costagliola,$^a$ Marco~Panero$^a$ and Arianna~Toniato$^{a,b}$\\
        \llap{$^a$}Department of Physics, University of Turin \& INFN, Turin\\
        Via Pietro Giuria 1, I-10125 Turin, Italy\\
        \llap{$^b$}CP$^3$-Origins \& Danish IAS, University of Southern Denmark\\
        Campusvej 55, 5230 Odense M., Denmark\\
        E-mail: \email{anada@to.infn.it}, \email{caselle@to.infn.it}, \email{costagli@to.infn.it}, \email{marco.panero@unito.it}, \email{toniato@cp3.sdu.dk}}

\abstract{%
{\it Preprint: CP3-Origins-2016-045 DNRF90}\\
\\
Jarzynski's equality is a well-known result in statistical mechanics, relating free-energy differences between equilibrium ensembles with fluctuations in the work performed during non-equilibrium transformations from one ensemble to the other. In this work, an extension of this relation to lattice gauge theory will be presented, along with numerical results for the $\mathbb{Z}_2$ gauge model in three dimensions and for the equation of state in $\SU(2)$ Yang-Mills theory in four dimensions. Then, further applications will be discussed, in particular for the Schr{\"o}dinger functional and for the study of QCD in strong magnetic fields.}

\FullConference{34th annual International Symposium on Lattice Field Theory\\
                 24-30 July 2016\\
                 University of Southampton, UK}

\begin{document}

\section{Introduction}

The computation of free energy differences in lattice gauge theories (LGTs) plays a crucial role in the study of a large set of physically interesting quantities and objects, such as 't Hooft loops, interfaces between center domains at non-zero temperature and many others.
The determination of the equation of state in QCD and QCD-like theories represents one of the most important examples: the pressure $p$ is indeed naturally related to the computation of the free energy density $f$, through the relation (valid in the thermodynamic limit) $p=-f$.

The numerical determination of free-energy differences in LGTs is a nontrivial computational challenge that motivates the search for new algorithms that can be easily implemented to study a large and diverse set of physical quantities.
Here we summarize our recent work~\cite{Caselle:2016wsw}, in which we presented a novel method for the computation of free energy differences based on the application to LGTs of Jarzynski's relation, a very well known result in statistical mechanics that was proved almost 20 years ago by Jarzynski~\cite{Jarzynski:1996ne, Jarzynski:1997ef}: it has been used since in a plethora of numerical studies in statistical mechanics and even verified experimentally (see, for instance, ref.~\cite{Liphardt:2002ei}).
We applied this method (briefly reviewed in section~\ref{sec:Jarzynski}) for the calculation of two very different quantities: the free energy of an interface in $\mathbb{Z}_2$ lattice gauge theory and the pressure of the $\SU(2)$ Yang-Mills gauge theory at non-zero temperature $T$. 
Results for these observables are presented in sections~\ref{sec:interface} and~\ref{sec:equation_of_state} and are discussed along with possible future applications in section~\ref{sec:conclusions}.

\section{Jarzynski's relation}
\label{sec:Jarzynski}

Jarzynski's equality relates the exponential statistical average of the work done on a system during a non-equilibrium process with the difference in free energy between the initial ($\Fin$) and the final ($\Ffin$) state of the system.
In the following, we denote the microscopic degrees of freedom of the system with $\phi$, so that the partition function $Z$ of a system with Hamiltonian $H$ (which depends also on a certain number of parameters and couplings) can be written as
\begin{equation}
\label{partition_function}
Z =\sum_{\phi} \exp \left( - \frac{H[\phi]}{T} \right).
\end{equation}
We introduce a set of parameters $\lambda$ (such as the couplings and/or the temperature $T$ itself), which are varied either continuously or discretely during a transformation from an initial value $\lambdain$, at which the system is in thermal equilibrium, to a final value $\lambdafin$.
The implementation for a Markov chain in a Monte Carlo simulation is straightforward: the process from the initial to the final state of the system is discretized into $N$ steps, each of which is characterized by a certain $\lambda_n$ (with $n=0,...N$).
After the $n$-th step the set of parameters takes the new value $\lambda_{n+1}$ which is used to update the old configuration $\phi_n$ of the system (obtained using $\lambda_{n}$) with the appropriate algorithm, thus driving it out of equilibrium and never letting it thermalize.

The quantity of interest is the total work done on the system when performing the transformation from $\lambdain$ to $\lambdafin$: it can be written as the sum over all the $N$ sub-intervals of the difference in the energy, i.e. in the Hamiltonian:
\begin{equation}
\label{discretized_exponential_work}
W(\lambdain,\lambdafin) = \sum_{n=0}^{N-1} H_{\lambda_{n+1}}\left[\phi_n\right] - H_{\lambda_n}\left[\phi_n\right]
\end{equation}
where $\lambda_0\equiv \lambdain$ and $\lambda_N \equiv \lambdafin$; the energies are evaluated using the same configuration $\phi_n$, before the system is updated with the new parameter $\lambda_{n+1}$.
Jarzynski's relation states the equality of the exponential average of the work over all possible isothermal transformations between the initial and final states and the exponential of the difference in free energy. It can be written as
\begin{equation}
\label{jarzynski_relation}
\left\langle \exp \left[ - \frac{W(\lambdain,\lambdafin)}{T} \right] \right\rangle = \exp \left( - \frac{\Ffin-\Fin}{T} \right)
\end{equation}
and a generalization for non-isothermal transformations has been introduced in ref.~\cite{Chatelain:2007ts}.
We remark that, in practice in a Monte Carlo simulation, the average is realized over a sufficiently large number of trajectories (which we denote with $\nr$) from the initial to the final state.

If we carry out a mapping of this relation from statistical mechanics to lattice gauge theory, we associate $H/T$ with the Euclidean action $S$ and transform the quantity appearing in eq.~\ref{discretized_exponential_work} into
\begin{equation}
\label{discretized_exponential_work_T}
\Delta S(\lambdain,\lambdafin) = \sum_{n=0}^{N-1} S_{\lambda_{n+1}}\left[\phi_n\right] - S_{\lambda_n}\left[\phi_n\right]
\end{equation}
and we can reexpress Jarzynski's relation using the ratio of the partition functions
\begin{equation}
\label{generalized_jarzynski_relation}
\left\langle \exp \left[ - \Delta S(\lambdain,\lambdafin) \right] \right\rangle = \frac{Z(\lambdafin)}{Z(\lambdain)}.
\end{equation}

There is an important remark to be made concerning the convergence of this method to the correct result, since a systematic uncertainty appears between the result obtained when performing a trajectory in a certain direction (``direct'' transformation) and in the opposite one (``reverse'' transformation). 
This discrepancy depends both on the discretization of the transformation into a finite number $N$ of steps and on the finite number $\nr$ of realizations: thus, the combination of $N$ and $\nr$ is chosen in order to meet the desired level of uncertainty (which can be set, for example, to be negligible with respect to the statistical error) while also minimizing the computational cost.
The determination of the optimal choice of $N$ and $\nr$ is a nontrivial problem which depends on the details of the system and/or the simulation; for a thorough discussion on the best practices to deal with this systematic uncertainty we refer to ref.~\cite{Pohorille:2010gp}.

\section{Benchmark study I: interface free energy in $\mathbb{Z}_2$ gauge model}
\label{sec:interface}

In the first part of this study we applied Jarzynski's relation to compute the free energy associated with the creation of an interface in the $\mathbb{Z}_2$ lattice gauge theory in 3 dimensions. 
The study of interfaces can be very insightful in high-energy physics: in particular they can be related to the world-sheet spanned by flux tubes in confining gauge theories and as such they can be analyzed both from the perspective of an effective theory and through numerical simulations on the lattice.

In the three-dimensional $\mathbb{Z}_2$ gauge model interfaces can be studied with extraordinary precision; here the degrees of freedom are $\mathbb{Z}_2$ variables defined on the links between nearest-neighbour sites of a cubic lattice. Remarkably, a confining phase exists for low values of the Wilson parameter $\betagauge$.
The Wilson action describing the dynamics of this model can be exactly rewritten using the Kramers-Wannier duality as the three-dimensional Ising model on the dual lattice, whose Hamiltonian reads
\begin{equation}
H = - \beta \sum_{x} \sum_{0 \le \mu \le 2} J_{x,\mu} \; s_x \; s_{x+a\hat{\mu}},
\end{equation}
where $s_x$ are $\pm1$ variables defined on the sites of the lattice, $J_{x,\mu}=\mp 1$ are (anti)ferromagnetic couplings from the site $x$ in the direction $\mu$, and $\beta=-\frac{1}{2} \ln \tanh \betagauge$.

An interface can be created by inducing a frustration on the system, i.e. by imposing the condition $J_{x,\mu}=-1$ for the couplings in a chosen direction and for a specific slice only of the lattice, while setting all the remaining ones to $+1$.
This is equivalent to imposing antiperiodic boundary conditions in one direction: the free energy of the interface (denoted as $F^{(1)}$) created this way can thus be defined as
\begin{equation}
\label{f1_defining_relation}
\frac{\Za}{\Zp} = N_0 \exp\left( - F^{(1)} \right)
\end{equation}
where $\Za$ and $\Zp$ are the partition functions of the system with antiperiodic and periodic boundary conditions respectively; the $N_0$ factor accounts for the possibility for the interface to be located anywhere in the $\mu=0$ direction.
An improved definition ($F^{(2)}$) which accounts for multiple interfaces has been introduced in \cite{Caselle:2007yc}.

The $\Za/\Zp$ ratio can be evaluated using Jarzynski's relation (eq.~\ref{jarzynski_relation}) by identifying the couplings $J_{x,\mu}$ used to create the frustration as the $\lambda$ parameters which are varied during the non-equilibrium transformation.
More specifically, we vary such couplings using
\begin{equation}
\label{J_evolution}
J_{x,\mu}(n)= 1 - \frac{2n}{N} \qquad {\rm for } \quad n=0,1,...,N
\end{equation}
which interpolates from $J=+1$ to $J=-1$ linearly; a similar implementation of Jarzynski's relation was used in recent works on the 2-dimensional Ising model \cite{condmat0602580, Chatelain:2007ts, Hijar2007}.

\begin{figure}
\centering
\includegraphics[width=0.65\textwidth]{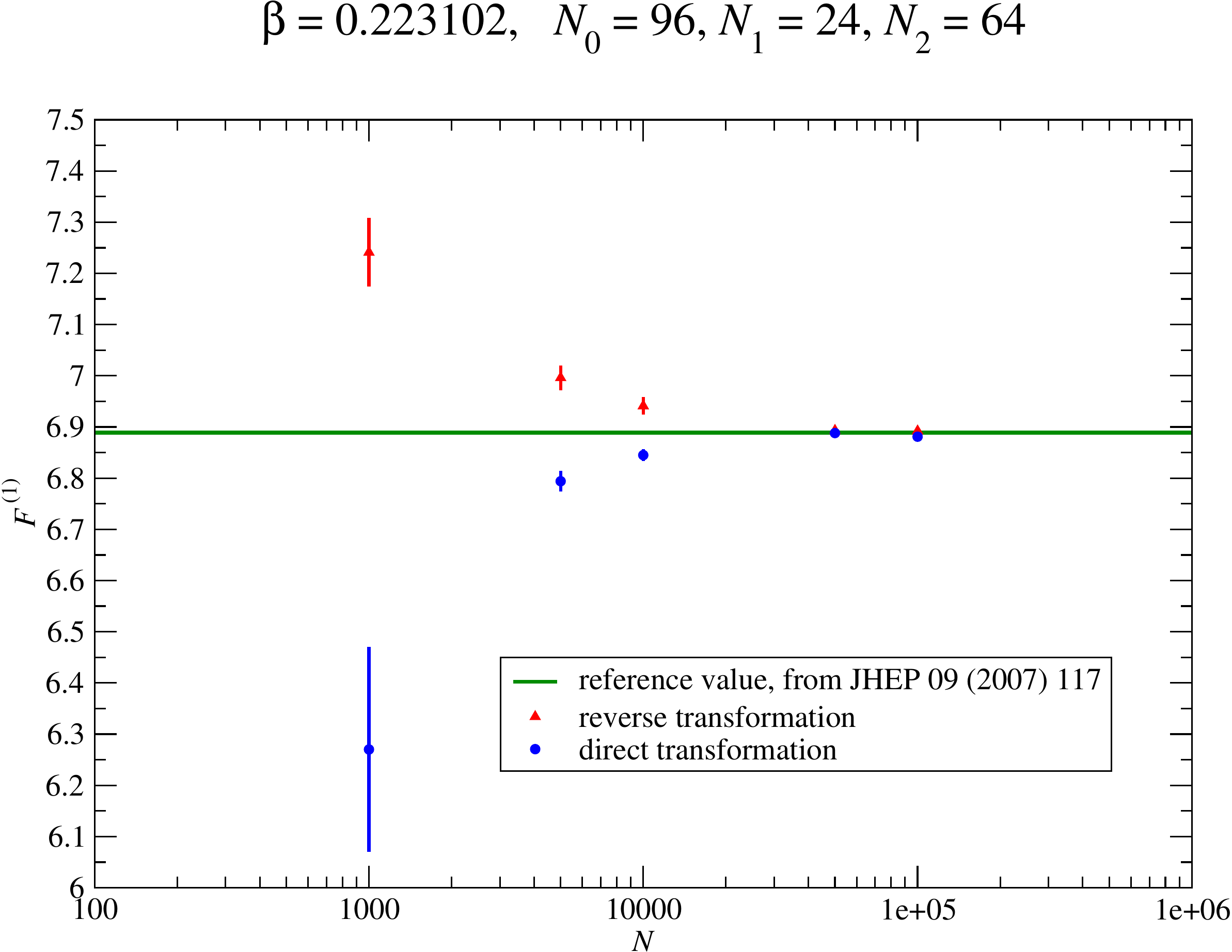}
\caption{Convergence of our results for the interface free energy obtained in direct and reverse transformations at $\beta=0.223102$ on a lattice of sizes $L_0=96a$, $L_1=24a$, $L_2=64a$ for increasing number of intermediate steps $N$.}
\label{fig:96_24_64}
\end{figure}

Results for the interface free energy obtained in Monte Carlo simulations using the ``direct'' (switching the couplings from $J=+1$ to $J=-1$) and the ``reverse'' (from $-1$ to $+1$) transformations are presented in fig.~\ref{fig:96_24_64}. 
They clearly converge on the same value of $F^{(1)}$ at fixed value of $\nr$ when the number of discretization steps is large enough; moreover, they show an excellent agreement with the results calculated in ref.~\cite{Caselle:2007yc} using thermodynamic integration.

After assessing the reliability of this method, we obtained high precision results at fixed $\beta$ for different values of the interface size $L$ using $N=10^6$ intermediate steps for each out-of-equilibrium transformation and carrying out the average of eq.~\ref{jarzynski_relation} over $\nr=10^3$ different trajectories.
These results can be compared with the analytical prediction of the effective string model that describes the transverse fluctuations of the interface at low energy.
In particular one can look at the difference between numerical results and the Nambu-Got\=o action prediction and examine its dependence on $L$, in order to understand the nature of the terms that do not arise from the Nambu-Got\=o low-energy expansion.
For the details of this analysis we refer to section 3 of ref.~\cite{Caselle:2016wsw}.

\section{Benchmark study II: pressure in $\SU(2)$ gauge theory}
\label{sec:equation_of_state}

In the second part of this study we focus on the determination of the equation of state in the confining phase of the $\SU(2)$ Yang--Mills theory in four spacetime dimensions.
We discretize it on a hypercubic lattice of spacing $a$ using Wilson's action;
the temperature is defined via $T=1/(aN_0)$, where $N_0$ is the extent of the periodic, compactified Euclidean-time dimension, while we take the lattice sizes in the three other directions to be equal ($N_1=N_2=N_3=N_s$) and sufficiently large to avoid finite-volume effects. 
Note that in order to control the temperature of the system, we used the relation between $a$ and $\betagauge$ determined in ref.~\cite{Caselle:2015tza}, so that we were able to change the temperature $T$ simply by varying $\betagauge$ at fixed $N_0$.
In this work we focus on the computation of the pressure $p$, which in the thermodynamic limit ($V \to \infty$) equals minus the free-energy density ($p= -f = -F/V$).

One of the most popular techniques use to calculate the pressure non-perturbatively on the lattice is the ``integral method'', introduced in ref.~\cite{Engels:1990vr}.
In a nutshell, the pressure as a function of the temperature $T$ is determined by integrating over $\betagauge$ plaquette expectation values $\langle U_{\Box}\rangle_T$ computed on lattices of size $N_0 \times N_s^3$ 
\begin{equation}
\label{lattice_pressure_im}
\frac{p(T)}{T^4} = 6 N_0^4 \int_{\betagauge^{(0)}}^{\betagauge^{(T)}} \dd \betagauge \left[ \langle U_{\Box}\rangle_{T} - \langle U_{\Box}\rangle_0 \right]
\end{equation}
where the lower integration limit $\betagauge^{(0)}$ corresponds to a temperature low enough at which the pressure is negligible.
Moreover, a quartic ultraviolet divergence has been removed by subtracting the value of the plaquette at $T=0$ (denoted as $\langle U_{\Box}\rangle_0$) computed on a symmetric lattice with size $\widetilde{N}^4$.

\begin{figure}
\centering
\includegraphics[width=0.65\textwidth]{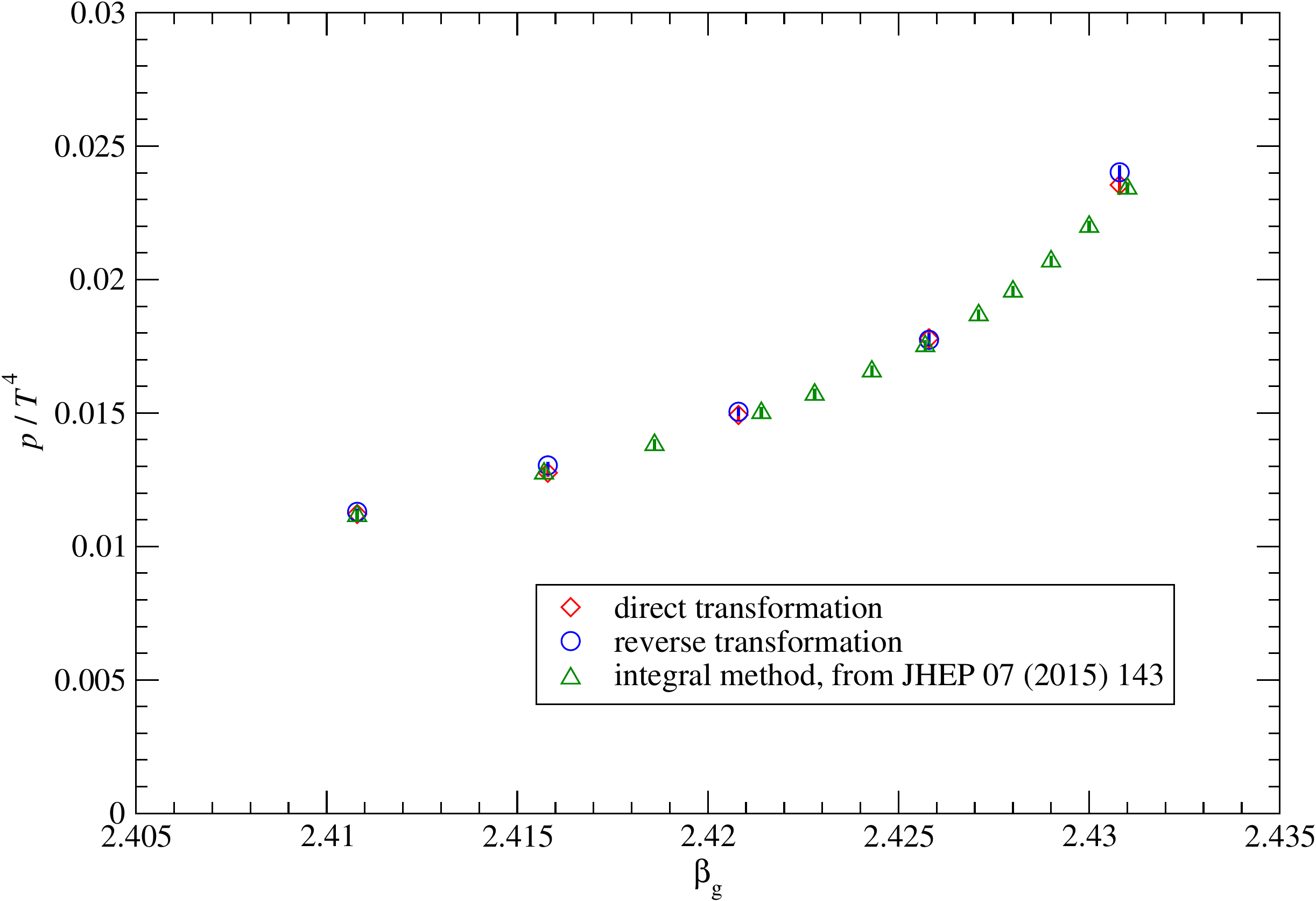}
\caption{Results for the pressure in the confining phase of $\SU(2)$ Yang-Mills theory, as a function of $\betagauge$ (which controls the temperature $T$), from simulations on lattices $N_s^3 \times N_0 = 72^3 \times 6$ (corresponding simulations at $T=0$ were carried out on $\widetilde{N}^4=40^4$ lattices).}
\label{fig:su2_pressure}
\end{figure}

Jarzynski's relation can be naturally extended to the computation of pressure differences by performing non-equilibrium transformations in Monte Carlo simulations on a $N_0 \times N_s^3$ lattice in which the role of the $\lambda$ parameter is taken by the Wilson parameter $\betagauge$.
The transformation starts at a certain value $\betagaugein$ (which corresponds to a certain temperature $T_0$) which is changed linearly after each update of the lattice variables until it reaches the final value $\betagaugefin$ (corresponding to the desired temperature $T$).
In this way eq.~\ref{generalized_jarzynski_relation} can be rewritten in order to compute differences in pressure:
\begin{equation}
\label{lattice_pressure_Jarzynski}
\frac{p(T)}{T^4} = \frac{p(T_0)}{T_0^4} + \left( \frac{N_0}{N_s} \right)^3 \ln \frac{ \langle \exp \left[ - \Delta S_{\SU(2)} (\betagaugein,\betagaugefin)_{N_0 \times N_s^3} \right] \rangle }{ \langle \exp \left[ - \Delta S_{\SU(2)} (\betagaugein,\betagaugefin)_{\widetilde{N}^4} \right] \rangle^{\gamma} }
\end{equation}
where $\Delta S_{\SU(2)}(\betagaugein,\betagaugefin)$ represents the total variation in the Wilson action (eq.~\ref{discretized_exponential_work_T}) in a transformation from $\betagaugein$ to $\betagaugefin$.
Like in eq.~\ref{lattice_pressure_im}, the divergence has been removed: the $T=0$ contribution is calculated performing the same transformation on a $\widetilde{N}^4$ lattice and then subtracting the corresponding total difference in the action; the exponent $\gamma = \left( N_s^3 \times N_0 \right) / \widetilde{N}^4$ is the ratio between the sizes of the two lattices.
Results for the confining phase have been calculated using eq.~\ref{lattice_pressure_Jarzynski} and reported in fig.~\ref{fig:su2_pressure}: very good convergence can be observed between ``direct'' and ``reverse'' transformations and moreover they show excellent agreement with older results obtained with the integral method from ref.~\cite{Caselle:2015tza}.
The transformations were performed independently from one value of $\betagauge$ to the next and they were discretized using either $N=1000$ or $N=2000$ intermediate steps; $\nr=30$ different realizations were carried out in order to compute the exponential average.
The implementation of Jarzynski's relation has proved to be more efficient than the integral method, since only a fraction of the configurations was needed to obtain results with comparable errors and to keep the systematic uncertainty under control.

\section{Discussion and future applications}
\label{sec:conclusions}

In this work we showed how an extension of Jarzynski's relation can be used to compute free-energy differences in lattice gauge theories. 
This novel method successfully reproduced results obtained with other techniques both for the interface free energy in the $\mathbb{Z}_2$ gauge model and the pressure in the $\SU(2)$ Yang-Mills theory.
Convergence to the correct result is achieved by increasing the number $\nr$ of realizations of the non-equilibrium transformation and/or the number of intermediate steps $N$ between initial and final state; under such conditions the efficiency of this method proved to be very competitive (and in many cases clearly superior) to that of other algorithms.

Since this novel method is very general and does not require strong assumptions, we envision a number of future applications of this relation with a particular attention to lattice gauge theories with dynamical fermions.
Specifically, we want to emphasize the possibility of applying Jarzynski's relation to studies involving the Schr\"odinger functional~\cite{Symanzik:1981wd, Luscher:1985iu} for the computation of the renormalized coupling $\bar{g}^2$: it could be used to compute changes in the effective action induced by changes in the parameters that specify the initial and final states on the boundaries of the lattice in the Euclidean time direction.
Another interesting application would be in the determination of the magnetic susceptibility of QCD in the presence of a strong background magnetic field $B$, which requires the computation of a free energy density difference between different values of $B$ (see for example ref.~\cite{Bonati:2013vba}).
We leave these and other potential applications to future studies.\\

\noindent{\bf Acknowledgements.}\\
The work of A.~T. is partially supported by the Danish National Research Foundation grant DNRF90.

\bibliography{jarzynski}

\end{document}